\documentstyle[12pt]{ioplppt}
\input psfig.sty

\newcommand{\vsp}{\vspace*{3mm}}
\newcommand{\bra}{\langle}
\newcommand{\ket}{\rangle}
\newcommand{\bd}{\begin{displaymath}}
\newcommand{\ed}{\end{displaymath}}
\newcommand{\be}{\begin{equation}}
\newcommand{\ee}{\end{equation}}

\def\ptl#1{\frac{\partial}{\partial #1}}
\def\bfm#1{\mbox{\boldmath $#1$}}
\def\dS{\mbox{\boldmath $S$}}

\def\bnul{\mbox{\boldmath $0$}}

\def\Z{{\cal Z}_\beta}
\def\tZ{\tilde{\cal Z}_{\tilde \beta}}

\begin{document}
\title{The XY Spin-Glass with Slow Dynamic Couplings}
\author{
G.~Jongen\ftnote{3}{E-mail: greetje.jongen@fys.kuleuven.ac.be}, 
D.~Boll\'e\ftnote{1}{E-mail: desire.bolle@fys.kuleuven.ac.be} and
A.C.C. Coolen\ftnote{2}{E-mail: tcoolen@mth.kcl.ac.uk}}
\address{\dag\S\ Instituut voor Theoretische Fysica,
            K.U.\ Leuven, B-3001 Leuven, Belgium }
\address{\ddag\
            Department of Mathematics, King's College London, 
            London WC2R 2LS, UK}
\begin{abstract}
\noindent
We investigate an XY spin-glass model in which both spins and couplings
evolve in time:   
the spins change rapidly according to Glauber-type rules, 
whereas the couplings evolve slowly with a dynamics
involving spin correlations and Gaussian disorder.
For large times the model can be solved using replica theory. In contrast
to the XY-model with static disordered couplings, solving the
present model requires two levels of replicas, one for the spins and one
for the couplings. Relevant order parameters are defined and a phase 
diagram is obtained upon making the replica-symmetric Ansatz. The
system exhibits two different spin-glass phases,
with distinct de~Almeida-Thouless lines, marking continuous
replica-symmetry breaking:  one describing  
freezing of the spins only, and one describing freezing of both spins
and couplings.  
\end{abstract}
\date{}


Recently various models with a coupled dynamics of fast Ising spins
and slow couplings have been studied (see e.g. \cite{CPS,C} and
references therein). 
In addition to physical motivations, such as understanding the
simultaneous learning and
retrieval in recurrent neural networks or the influence of slow atomic
diffusion processes in disordered magnetic systems, 
there is a  more theoretical interest in such models in that
they generate the replica formalism for a {\em finite} number of
replicas $n$. Moreover, the replica number is found to have a physical
meaning as the ratio of two temperatures (characterizing the 
stochasticity in the 
spin dynamics and the coupling dynamics, respectively). 
In this letter we extend the methods and results obtained for Ising 
spin models to a classical XY spin-glass with dynamic couplings, 
whose spin variables are physically more realistic than Ising ones. 
In addition, the XY model is closely related to neural network models of
coupled oscillators, which provide a phenomenological description of 
neuronal firing synchronization in brain tissue. We solve our model
upon making the replica-symmetric Ansatz, and calculate the 
de Almeida-Thouless (AT) lines \cite{AT} 
(of which here there are two types), where continuous transitions
occur to phases with broken replica symmetry. In doing so we
also improve the calculations of \cite{PS}.  
As in the Ising case 
we find two qualitatively different types of spin-glass phases. In  one
spin-glass phase the spins do freeze in random directions, but on the
time-scales of the coupling dynamics these `frozen directions'
change. In the second spin-glass phase the spins as well as the
couplings freeze, such that even on the large time-scales the `frozen
directions' of the spins remain stationary. 
\vsp

We choose a system of $N$ classical two-component spin variables
$\dS_i=(\cos\theta_i,\sin\theta_i)$, $i=1\dots N$, and symmetric  
exchange interactions (or couplings) $J_{ij}$,  with a
Glauber-type dynamics such that for {\em stationary} choices of the
couplings the microscopic spin probability density would evolve
towards a Boltzmann distribution, with the standard Hamiltonian   
$H(\{\dS_i\},\{J_{ij}\})=-\sum_{k<\ell}J_{k\ell}\dS_k\cdot\dS_\ell$ 
and with inverse temperature $\beta=T^{-1}$. 
The couplings $J_{ij}$ are taken to be of infinite range. They
will now themselves be allowed to evolve in time in a stochastic
manner, partially in response to 
the states of the spins 
and to externally imposed biases. 
However, we  assume that the spin dynamics is very fast
compared to that of the couplings, such that on the time-scales of
the couplings, the spins are effectively in equilibrium (i.e. we take 
the adiabatic limit). 
For the dynamics of the couplings the following Langevin form is 
proposed (which is the natural adaptation to XY spins of the  
choices originally made in \cite{CPS,PS} for Ising spins):
\begin{equation}
        \frac{d}{dt}J_{ij}=
                \frac{\left<\dS_i\cdot\dS_j\right>+K_{ij}}{N}
                -\mu J_{ij}
                +\frac{\eta_{ij}(t)}{N^{1/2}} 
~~~~~~~~~~~~ i<j=1\dots N \, .
        \label{def:evcouplings}
\end{equation}
The term $\left<\dS_i\cdot\dS_j\right>$, representing local spin
correlations associated with the coupling $J_{ij}$, is a thermodynamic
average over the Boltzmann distribution of the spins, given the
instantaneous couplings $\{J_{k\ell}\}$. 
External biases $K_{ij}\!=\!\mu N B_{ij}$ serve to steer the weights to
some preferred values (note: this notation follows that of \cite{PS}). 
The $B_{ij}$ are choosen to be quenched random variables, drawn 
independently from a Gaussian probability distribution with mean
$B_0/N$ and variance $\tilde{B}/N$.  
The decay term $\mu J_{ij}$ in (\ref{def:evcouplings}) 
is added to limit the magnitude of the couplings.
Finally, the terms $\eta_{ij}(t)$ represent Gaussian white noise 
contributions,  of zero mean
and covariance $\bra \eta _{ij}(t)\eta _{kl}(t')\ket =
         2{\tilde T}~\delta _{ik}\delta _{jl}\delta (t-t')$, 
with associated temperature $\tilde T = {\tilde \beta}^{-1}$. 
Factors of $N$ are introduced in order to ensure non-trivial behaviour in
the thermodynamic limit $N \rightarrow \infty$.

The three independent global symmetries of our model, which can
be expressed efficiently in terms of the Pauli spin matrices
$\sigma_x$ and $\sigma_z$, are the following:
\be
\begin{array}{llll}
{\rm inversion~of~both~spin~axes:}&& \dS_i\to -\dS_i & {\rm
for~all}~i\\
{\rm inversion~of~one~spin~axis:}&& \dS_i\to \sigma_z\dS_i & {\rm
for~all}~i\\
{\rm permutation~of~spin~axes:}&& \dS_i\to \sigma_x\dS_i & {\rm
for~all}~i \, .
\end{array} 
\label{eq:symmetries}
\ee
Upon using algebraic relations such as
$\sigma_x\sigma_z\sigma_x=-\sigma_z$ 
and $\sigma_z\sigma_x\sigma_z=-\sigma_x$ we see that in the
high $T$ (ergodic) regime these three global symmetries generate the following
local identities, respectively:
\be
\bra \dS_i\ket=\bnul,~~~~~~~~
\bra \dS_i\cdot\sigma_x\dS_j\ket=0,~~~~~~~~
\bra \dS_i\cdot\sigma_z\dS_j\ket=0 \, .
\label{eq:identities}
\ee

The equilibrium solution of the probability density associated with the
stochastic equation (\ref{def:evcouplings}) for the couplings follows
from the fact that (\ref{def:evcouplings}) 
is conservative, i.e. that it can be written as
\be
\frac{d}{dt}J_{ij}=-\frac{1}{N}\ptl{J_{ij}} \tilde H(\{J_{ij}\})
+\frac{\eta_{ij}(t)}{N^{1/2}} 
\label{eq:conservative}
\ee
with the following effective Hamiltionian for the couplings:
\begin{equation}
        \tilde H(\{J_{ij}\})=-\frac1\beta\log\Z(\{J_{ij}\})
                +\frac12\mu N\sum_{k<\ell}J_{k\ell}^2
                -\mu N\sum_{k<\ell}B_{k\ell}J_{k\ell}\,.
\label{eq:couplinghamiltonian}
\end{equation}
In this expression $\Z(\{J_{ij}\})={\rm Tr}_{\{\dS_i\}}\exp[\beta
\sum_{k<\ell}J_{k\ell}\dS_k\cdot\dS_\ell]$ is the partition function
of the XY spins  with instantaneous couplings $\{J_{ij}\}$. 
Thus the stationary probability density for
the couplings is also of a Boltzmann form, with the Hamiltonian
(\ref{eq:couplinghamiltonian}), 
and the thermodynamics of the slow system (the couplings) are  
generated by the partition function $\tZ=\int \prod_{k<\ell}
dJ_{k\ell}~\exp[-\tilde{\beta}\tilde{H}(\{J_{ij}\})]$, leading to 
(modulo irrelevant prefactors):
\begin{equation}
\hspace*{-5mm}
        \tZ=\int \prod_{k<\ell} dJ_{k\ell}
                \left[\Z(\{J_{ij}\}) \right]^n
\exp\left[\mu \tilde\beta N  \sum_{k<\ell}B_{k\ell}J_{k\ell}
-\frac{1}{2}~\mu \tilde \beta N\sum_{k<\ell}J_{k\ell}^2 \right] \, .
       \label{eq:partition function}
\end{equation}
Finally, we define the disorder-averaged free energy per site
$\tilde f=-(\tilde\beta N)^{-1}\bra \log\tZ\ket_B$, in which 
 $\bra \cdot \ket_B$ is an average over the $\{B_{ij}\}$.  
In contrast to standard systems with frozen disorder, the replica
number $n$ is here given by the ratio $n=\tilde\beta/\beta$, and can
take any real non-negative value. The limit
$n\rightarrow0$ corresponds to a situation in which the coupling 
dynamics is driven by the Gaussian white noise, rather than by 
spin correlations; in the limit $n\rightarrow\infty$ the influence of
spin correlations dominates. 

We carry out the disorder average  using the identity 
$\log \tZ=\lim_{r\rightarrow0}r^{-1}[\tZ^r\!-1]$, evaluating the 
latter by analytic continuation from integer $r$. Our system with
partition function $\tZ$ is thus replicated $r$ times; 
we label each replica by a Roman index.
Each of the $r$ functions $\tZ$, in turn, is given by 
(\ref{eq:partition function}), and involves 
$\Z(\{J_{ij}\})^n$ which is replaced by the product of
$n$ further replicas, labeled by Greek indices.
For non-integer $n$, again analytic continuation is made
from integer $n$. Performing the
disorder average in $\tilde{f}$ results in an expression involving
$nr$ coupled replicas of the original system: $\{\dS_i\}\to
\{\dS_{ia}^\alpha\}$, with $\alpha=1\ldots n$ and $a=1\ldots r$. 
For $N\to\infty$ this expression can be evaluated in the familiar 
fashion of replica mean-field theory \cite{MPV}, by saddle-point
integration. This procedure induces the following order parameters:
\bd
 \bfm{m}_a^\alpha
=\frac1N\sum_{i}
\left<~\overline{\left<
\dS_{ia}^{\alpha}\right>}~\right>_{\!B}
~~~~~~~~~~~~~
 q_{ab}^{\alpha\beta}=\frac1N\sum_{i}
        \left<~\overline{\left<
       \dS_{ia}^{\alpha}\cdot\dS_{ib}^{\beta}\right>}~\right>_{\!B}
\ed
\bd
u_{ab}^{\alpha\beta}=\frac1N\sum_{i}\left<~\overline{\left<
\dS_{ia}^{\alpha}\cdot
\sigma_x \dS_{ib}^{\beta}
        \right>}~\right>_{\!B} 
~~~~~
v_{ab}^{\alpha\beta}=\frac1N\sum_{i}\left<~\overline{\left<
\dS_{ia}^{\alpha}\cdot
\sigma_z \dS_{ib}^{\beta}
        \right>}~\right>_{\!B}  \, .
\ed
The horizontal bar denotes thermal averaging over the 
coupling dynamics with fixed biases $\{B_{ij}\}$. Comparison with 
(\ref{eq:identities}) shows that the order parameters
$\bfm{m}_a^\alpha$, $u_{ab}^{\alpha\beta}$ and $v_{ab}^{\alpha\beta}$ 
measure the breaking of the global symmetries (\ref{eq:symmetries}).  
For simplicity we choose $B_0=0$. We make the usual
assumption that, in the absence of global symmetry-breaking forces, 
phase transitions can lead to at most {\em local} violation of the
identities (\ref{eq:identities}). Thus the latter will remain valid 
if averaged over all sites, at any temperature, which implies
that $\bfm{m}_a^\alpha \!=\! \bnul$ and 
$u_{ab}^{\alpha\beta}\!=\!v_{ab}^{\alpha\beta}\!=\!0$.  
The spin-glass order parameters $q_{ab}^{\alpha\beta}$, 
on the other hand, are not related to simple global symmetries, and
serve to characterize the various phases.  

The final stage of the calculation is to make the replica symmetry
(RS) Ansatz. 
Since observables with identical Roman indices refer to system copies with 
identical couplings, whereas observables with identical Roman indices {\em
and} identical Greek indices refer to system copies with 
identical couplings {\em and} identical spins,  in the present
model the RS Ansatz takes the form 
$q_{ab}^{\alpha\beta}=\delta_{ab}\left\{\delta_{\alpha\beta}+q_1[1-
\delta_{\alpha\beta}]\right\}+q_{0}[1-\delta_{ab}]$
(note: $\dS_{a}^\alpha\cdot\dS_{a}^\alpha=1$). 
The remaining two order parameters are determined as the solutions of
the following coupled saddle-point equations:
\be
\hspace*{-5mm}
q_0=\int\!dx~P(x)\left\{
\frac{\int\!dz~P(z)\left[ I_0(z\Xi)\right]^{n-1}
I_1(z\Xi)~
I_1(z x\beta \Xi^{-1}\sqrt{\frac{1}{2}\tilde{B}q_0})}
{\int\!dz~P(z)\left[ I_0(z\Xi)\right]^{n}
I_0(z x\beta \Xi^{-1}\sqrt{\frac{1}{2}\tilde{B}q_0})}
\right\}^2
\label{eq:saddle1}
\ee
\be
\hspace*{-5mm}
q_1=\int\!dx~P(x)\left\{
\frac{\int\!dz~P(z)\left[ I_0(z\Xi)\right]^{n-2}
\left[I_1(z\Xi)\right]^{2}
I_0(z x\beta \Xi^{-1}\sqrt{\frac{1}{2}\tilde{B}q_0})}
{\int\!dz~P(z)\left[ I_0(z\Xi)\right]^{n}
I_0(z x\beta \Xi^{-1}\sqrt{\frac{1}{2}\tilde{B}q_0})}
\right\}
\label{eq:saddle2}
\ee
with the two short-hands $\tilde J
\!=\!1/\mu\tilde\beta$, $\Xi\!=\!
\beta\sqrt{\frac{1}{2}(\tilde{J}\!+\!\tilde{B})q_1\!- 
\!\frac{1}{2}\tilde{B}q_0}$, 
with $P(x)\!=\!x e^{-\frac{1}{2}x^2}\theta[x]$, and where the 
functions $I_n(x)$ are the Modified Bessel functions \cite{Abramowitz}. 
Their physical meaning is given by 
\be
q_{0}=\frac{1}{N}\sum_i\left<~\overline{\left<\dS_i\right>}^2~\right>_B
~~~~~~~~~~~~
q_1=\frac{1}{N}\sum_i\left<~\overline{\left<\dS_i\right>^2}~\right>_B
     \, .
\label{eq:meaning}
\ee
It is clear that $0\leq q_{0}\leq q_1\leq 1$.
\vsp

\begin{figure}[hb]
\centerline{
\psfig{figure=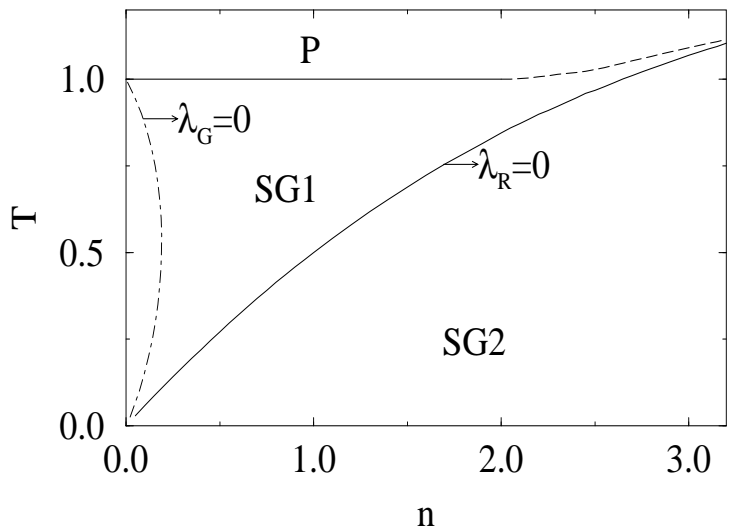,height=6.5cm,width=9.5cm,angle=270
}
\hspace*{10mm}
}
\caption{Phase diagram of the XY spin-glass with slow dynamic couplings,
drawn in the $n$-$T$ plane; for $B_0=0$, $\tilde B=1$ and $\tilde
J=3$. P: paramagnetic phase, $q_1=q_0=0$. 
SG1: first spin-glass phase, $q_1>0$ and $q_0=0$ (freezing on spin
time-scales only).   
SG2: second spin-glass phase, $q_1>0$ and $q_0>0$ (freezing on all 
time-scales). 
AT lines: $\lambda_R=0$ (Roman replicon), $\lambda_G=0$ (Greek replicon).   
}
\label{fig:ntxy}
\end{figure}

\begin{figure}[b]
\centerline{
\psfig{figure=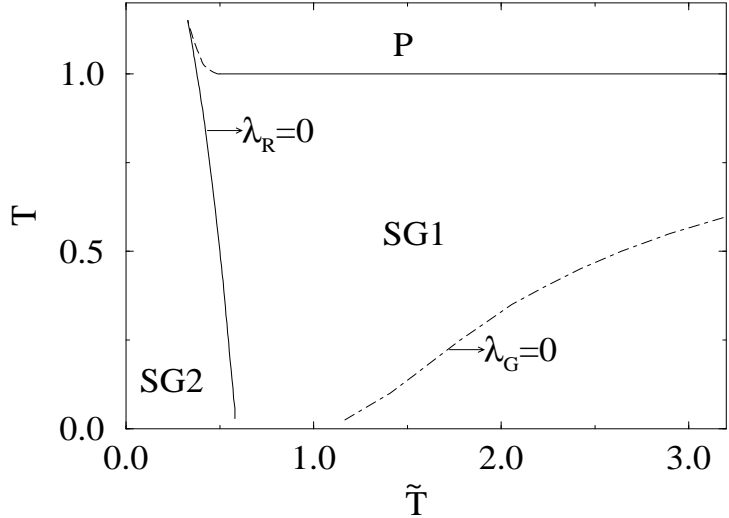,height=6.5cm,width=9.5cm,angle=270
}
\hspace*{10mm}
}
\caption{Phase diagram of the XY spin-glass with slow dynamic couplings,
drawn in the $\tilde{T}$-$T$ plane; for $B_0=0$, $\tilde B=1$ and $\tilde
J=3$. Further notation as in Figure~\protect\ref{fig:ntxy}.}
\label{fig:ttildetxy}
\end{figure}

\begin{figure}[b]
\centerline{
\psfig{figure=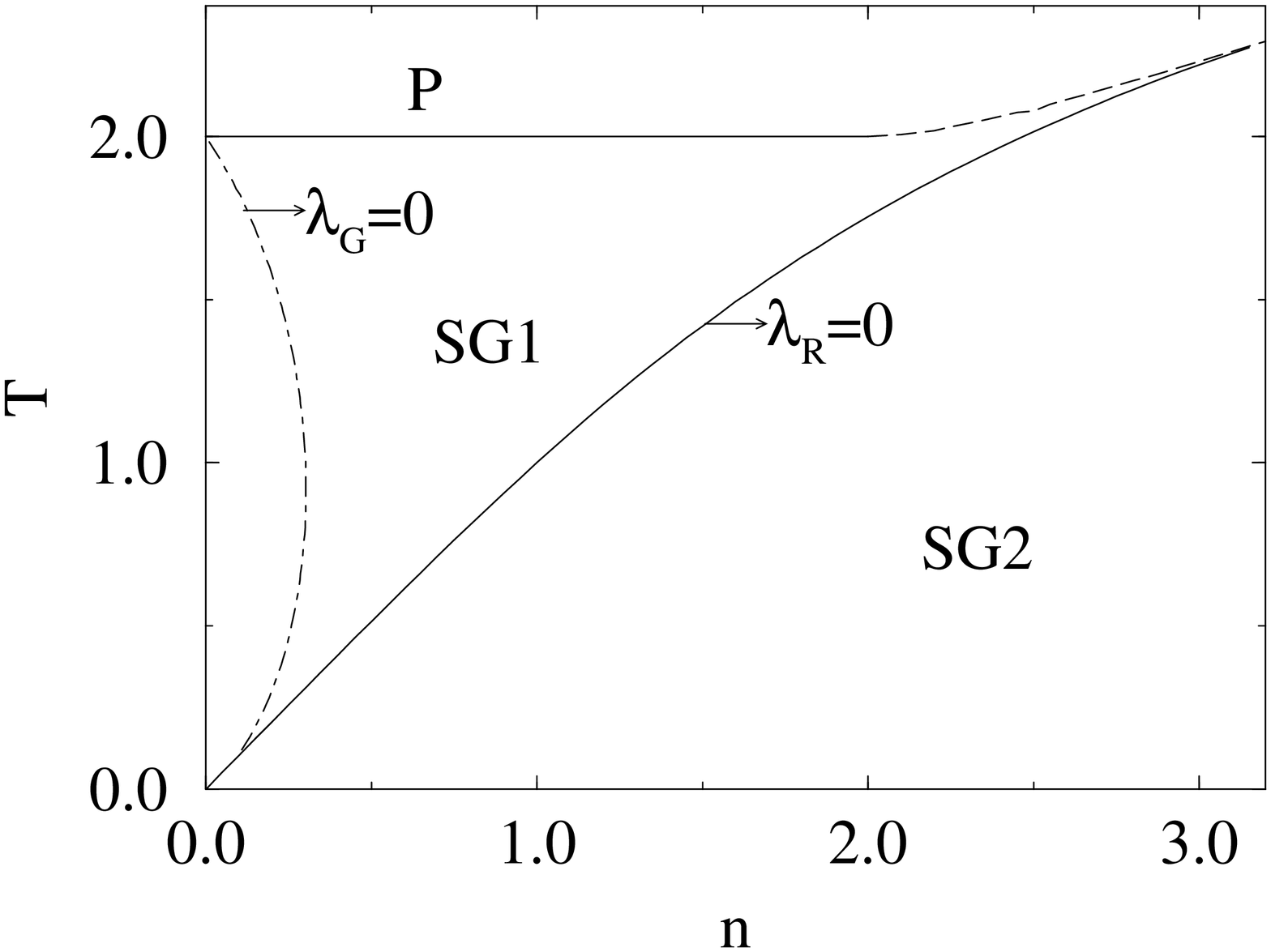,height=6.5cm,width=9.5cm,angle=270}
\hspace*{10mm}
}
\caption{Phase diagram of the Ising spin-glass with slow dynamic
couplings [1],
drawn in the $n$-$T$ plane; for $B_0=0$, $\tilde B=1$ and $\tilde
J=3$. Further notation as in Figure~\protect\ref{fig:ntxy}.
}  
\label{fig:SK} 
\end{figure}
We have studied the fixed-point equations
(\ref{eq:saddle1},\ref{eq:saddle2})  after having first eliminated 
the parameter redundancy by putting  $\tilde B=1$ and $\tilde J=3$, 
which resulted in the phase diagram 
in the $n$-$T$ plane as shown in figure \ref{fig:ntxy}.  
In addition to a paramagnetic phase (P), where $q_0=q_1=0$,  
one finds two distinct spin-glass phases: SG1, where $q_{1}>0$ but
$q_0=0$, and  SG2, where both $q_1>0$ and $q_0>0$. 
The SG1 phase describes freezing of the spins on the fast time-scales
only (where spin equilibration occurs); on the large time-scales, where
coupling equilibration occurs, one finds that, due to the slow motion of
the couplings, the frozen spin directions continually change. 
In the SG2 phase, on the other hand, both spins and couplings freeze,
with the net result that even on the large time-scales the frozen spin
directions are `pinned'. The SG1-SG2 transition is always second order. 
The transition SG1-P is second order for
$n<2$ (in which case its location is given by 
$\tilde B + \tilde J=4T^2$), but first order for $n>2$. 
When $n$ further increases to $n>3.5$, the SG1 phase disappears, and the
system exhibits a first order transtion from P to SG2. 

Additional transitions occur, corresponding to a continuous breaking of
replica symmetry. The stability of the RS solutions is, as always, 
expressed in terms of the eigenvalues of the
matrix of second derivatives of quadratic fluctuations at the saddle-point
\cite{AT}. We have calculated all eigenvalues and their
multiplicity following \cite{PS,AT} (full details will be published 
elsewhere \cite{fullpaper}). 
We find two replicon eigenvalues: $\lambda_G$, associated with the Greek
replicas and $\lambda_R$, associated with the Roman replicas. We have
drawn the corresponding AT-lines, where $\lambda_G$ and $\lambda_R$ are 
zero, respectively, in the phase diagram as dashed-dotted lines. 
At low spin temperature $T$  replica symmetry is broken with respect to
the Greek replicas ($\lambda_G=0$), whereas at low coupling
temperature $\tilde{T}$ replica symmetry is broken with respect to
the Roman replicas ($\lambda_R=0$). This is illustrated more clearly by
drawing the phase diagram in the $\tilde{T}$-$T$ plane, as in figure 
\ref{fig:ttildetxy}. 
This second figure also shows that there is no re-entrance from SG1 to 
RSB, when $T$ is varied for fixed $\tilde{T}$. 

Qualitatively the phase diagram of the present system is very similar to 
that of the Ising spin-glass with dynamic couplings as studied in 
\cite{PS}, see figure \ref{fig:SK}, including the behaviour of both the 
Greek AT-line $\lambda_G=0$ and the Roman AT-line $\lambda_R=0$. 
Here our results differ from, and improve upon, those of \cite{PS} 
(which is why we present figure \ref{fig:SK}, rather than just refer to
\cite{PS}). The set of eigenvalues given in \cite{PS} turn out to   
satisfy only part of the relevant orthogonality conditions used in their
calculation. The replica symmetric solution in SG1 is always stable with
respect to the Roman replicas. In fact, we can show analytically that 
the Roman AT-line coincides with the SG1-SG2 transition line.
Our model, with dynamics on two different time-scales, is reminiscent of 
a simple XY model with one step replica-symmetry breaking (1RSB), and 
our eigenvalues formally resemble e.g. those describing the stability of 
the 1RSB solution in the perceptron model of \cite{WS}. 
\vsp

In conclusion, we have solved a classical XY spin-glass model in which 
both the spins and their couplings evolve stochastically, according to
coupled equations, but on different time-scales. 
The solution of our model in RS Ansatz is mathematically similar to
that of the XY model with static couplings, but with one step RSB.
Qualitatively, the phase diagram, which exhibits two different 
spin-glass types and both first and second order transitions, resembles 
closely that of the Ising spin-glass with dynamic couplings, provided
appropriate adjustment of the calculation of the AT-line in \cite{PS} 
is carried out. Our calculation shows how the methods used for solving 
the Ising case can be easily adapted to more complicated spin types, and
illustrates the robustness of the structure and peculiarities of  
phase diagrams describing the behaviour of large 
spin systems with dynamic couplings. 

\subsection*{Acknowledgments}

We would like to thank R.W. Penney and D. Sherrington for helpful
discussions. 
This work has been supported in part by the Research Fund of the
K.U. Leuven  (grant OT/94/9). DB and ACCC would like to thank the Fund 
for Scientific Research-Flanders (Belgium) and the British Council for
financial support. 
\vsp\vsp

\end{document}